\documentclass{elsarticle}

\pdfoutput=1

\usepackage{url}
\usepackage{here}
\usepackage{amsmath}
\usepackage{listings}
\usepackage{subfig}

\makeatletter
\def\ps@pprintTitle{%
 \let\@oddhead\@empty
 \let\@evenhead\@empty
 \def\@oddfoot{}%
 \let\@evenfoot\@oddfoot}
\makeatother

\begin{document}

\begin{frontmatter}
\date{}

\title{Population-based de novo molecule generation, using grammatical evolution}

\author[1]{Naruki~Yoshikawa}
\author[1]{Kei~Terayama}
\author[2]{Teruki~Honma}
\author[3]{Kenta~Oono}
\author[1,4,5]{Koji~Tsuda\corref{cor}}
\cortext[cor]{Correspondence: tsuda@k.u-tokyo.ac.jp}
\address[1]{Department of Computational Biology and Medical Sciences, Graduate School of Frontier Sciences, The University of Tokyo, Kashiwa, Japan}
\address[2]{RIKEN Systems and Structural Biology Center, Yokohama, Japan}
\address[3]{Preferred Networks, Inc., Tokyo, Japan}
\address[4]{National Institute for Materials Science, Tsukuba, Japan}
\address[5]{RIKEN Center for Advanced Intelligence Project, Tokyo, Japan}

\begin{abstract}
Automatic design with machine learning
and molecular simulations has shown a remarkable ability
to generate new and promising drug candidates.
Current models, however, still have problems in simulation concurrency
and molecular diversity.
Most methods generate one molecule at a time and do not allow multiple
simulators to run simultaneously.
Additionally, better molecular diversity could boost the success rate in
the subsequent drug discovery process.
We propose a new population-based approach
using grammatical evolution named ChemGE.
In our method, a large population of molecules are updated concurrently
and evaluated by multiple simulators in parallel.
In docking experiments with thymidine kinase,
ChemGE succeeded in generating hundreds of high-affinity molecules
whose diversity is better than that of known binding molecules in DUD-E.
\end{abstract}
\end{frontmatter}

\section{Introduction}
Designing new molecules with desirable properties is an important task
for pharmaceutical science and materials science, with the improvement of binding affinity and ADMET profile, a priority for hit-to-lead and subsequent optimization stages of drug discovery.
Fragment-based methods, such as RECAP~\cite{lewell1998}, generate molecules
by linking known fragments, but these methods have problems in
structural diversity and patentability.
Recently, several \textit{de novo} molecule generation methods that do not
require fragments have been proposed.
They formulate the molecule generation problem
as a black-box optimization of SMILES
strings~\cite{weininger1988} and
solve it with deep neural networks and molecular simulations.
Recent approaches include
(1) Bayesian optimization, over a continuous space, of variational autoencoders~\cite{gomez2016, kusner2017},
(2) optimization of recurrent neural network through fine-tuning or reinforcement learning~\cite{guimaraes2017, segler2017, popova2017},
(3) sequential Monte Carlo search over a language model of SMILES~\cite{ikebata2017},
and (4) Monte Carlo tree search guided by recurrent neural
network~\cite{yang2017}.

Although these methods achieved great results in finding drug candidates,
most still have problems in simulation concurrency
and molecular diversity.
Few, often no more than one, molecules are generated at a
time, making it difficult to parallelize their subsequent evaluation through molecular simulation.
The lack of diversity in generated molecules is also a problem~\cite{benhenda2017}, due mainly to the way most methods are designed to find an optimal molecule based on an \textit{a priori} defined score: drug discovery is a stage-wise process and it is impossible
to design a perfect score reflecting all aspects.
Improving molecular diversity is necessary to increase the chance of
survival, down the drug discovery pipeline.

\begin{figure*}[ht]
    \centering
    \includegraphics[width=0.9\textwidth]{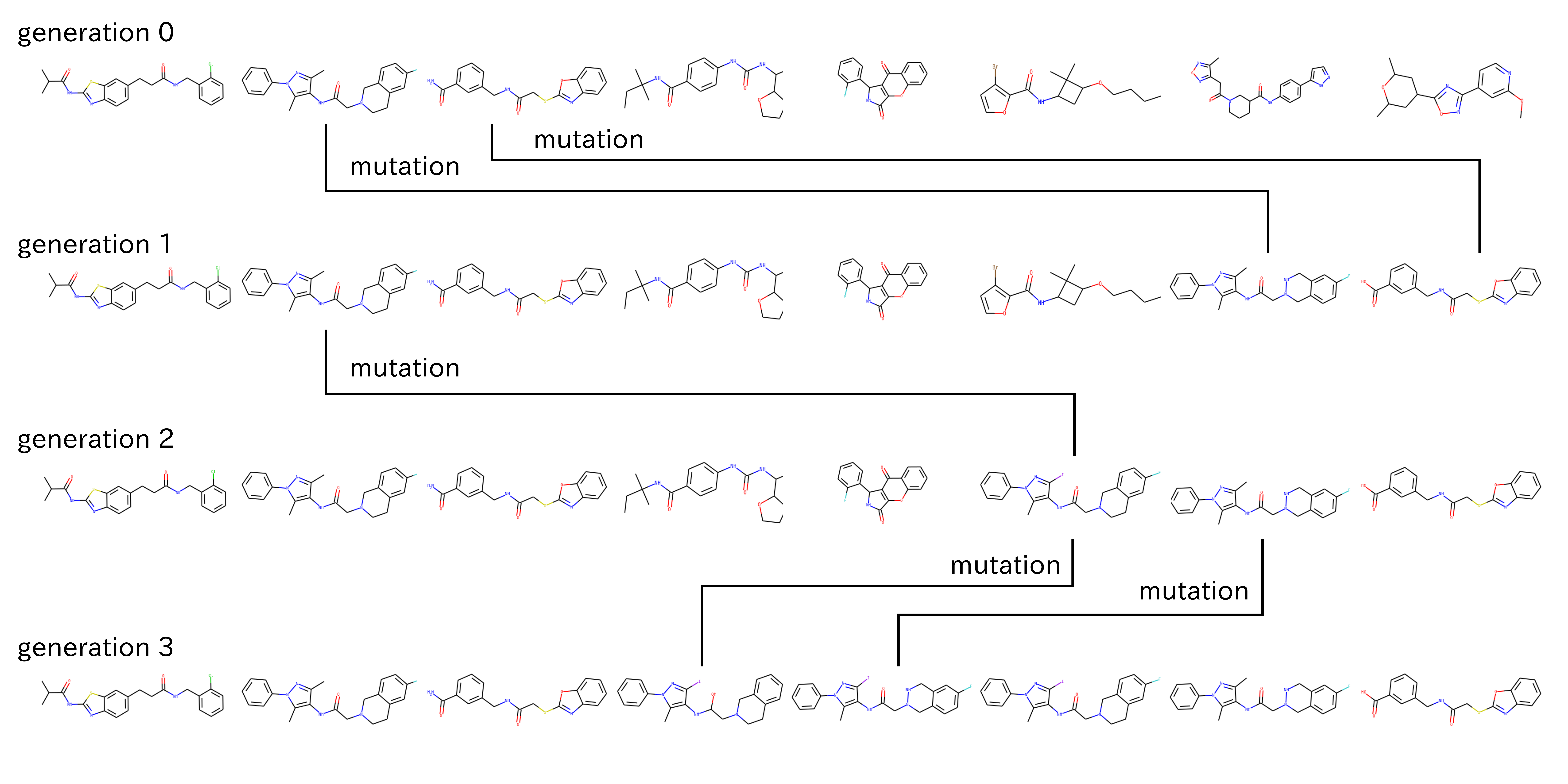}
   \caption{\textbf{Population-based molecular design with ChemGE.}
   A population of molecules are updated to achieve better fitness. When we aim to generate molecules docking to a target protein, the fitness is set to the affinity score of a docking simulator.
   A new generation of molecules is made by choosing some molecules and modifying them (i.e., mutation).}
   \label{fig:illust_evolution}
\end{figure*}

In this paper, we propose a new population-based approach named ChemGE,
which uses grammatical evolution~\cite{dempsey2009}
to optimize a population of molecules (Figure~\ref{fig:illust_evolution}).
Grammatical evolution is a population-based optimization technique
to optimize a population of strings that follow a context-free grammar.
Such population-based evolutionary methods have been
regaining popularity for solving black-box optimization problems, such as
hyperparameter optimization and neural network design (e.g.,\cite{jaderberg2017}).
Their main advantage is inherent concurrency: ChemGE allows many simulations to run simultaneously and can easily be adapted for parallel computation.
In addition, mutation operations in evolution
ensure large diversity throughout the optimization process.
In benchmarking experiments using a druglikeness score,
we confirmed that ChemGE is able to find many more molecules than deep
learning-based methods using identical computational resources.
ChemGE was also used to find novel molecules docking to thymidine
kinase using rDock~\cite{rdock}.
Employing 32 cores in parallel, ChemGE generated 349 molecules
whose docking scores are better than the best molecule in
DUD-E~\cite{dud-e} in 26 hours. The generated molecules were
highly diverse, and dissimilar to known active ones in DUD-E.

\section{Method}
\subsection{Context-free grammar of SMILES}
This section reviews SMILES and its context-free grammar.
A context-free grammar $G$ is defined
as a 4-tuple: $G=(V, \Sigma, R, S)$,
where $V$ is a finite set of non-terminal symbols,
$\Sigma$ is a finite set of terminal symbols,
$R$ is a finite set of production rules,
and $S$ is a special non-terminal symbol called the start symbol.
A production rule defines a transformation from
a non-terminal symbol $V$ to $(V \cup \Sigma)^*$,
where the asterisk denotes a Kleene star.
To generate a string, a sequence of production rules is
applied to non-terminal symbols until no non-terminal symbol remains.

SMILES (Simplified Molecular Input Line Entry System)\cite{weininger1988} is a notation of
chemical structure using ASCII strings.
OpenSMILES~\cite{james2007} specifies the context-free grammar of SMILES.
In this paper, we used a subset of the grammar shown in Figure~\ref{fig:grammar}.
Note that following the grammar alone is not sufficient for generating valid molecules.
For example, \verb/C#O/ indicates a molecule where a carbon and oxygen atoms are connected by a triple bond, but such a molecule does not exist.
A string is called a valid SMILES, if the string represents a molecule.
One can use, e.g., RDKit~\cite{rdkit} to check the validity of a generated string.

\begin{figure*}[h!]
\begin{tabular}{c}
\begin{lstlisting}[basicstyle=\ttfamily\tiny]
<smiles> ::= <chain>
<atom> ::= <bracket_atom> | <aliphatic_organic> | <aromatic_organic>
<aliphatic_organic> ::= "B" | "C" | "N" | "O" | "S" | "P" | "F" | "I" | "Cl" | "Br"
<aromatic_organic> ::= "c" | "n" | "o" | "s"
<bracket_atom> ::= "[" <BAI> "]"
<BAI> ::= <isotope> <symbol> <BAC> | <symbol> <BAC> | <isotope> <symbol> | <symbol>
<BAC> ::= <chiral> <BAH> | <BAH> | <chiral>
<BAH> ::= <hcount> <BACH> | <BACH> | <hcount>
<BACH> ::= <charge> <class> | <charge> | <class>
<symbol> ::= <aliphatic_organic> | <aromatic_organic>
<isotope> ::= <DIGIT> | <DIGIT> <DIGIT> | <DIGIT> <DIGIT> <DIGIT>
<DIGIT> ::= "1" | "2" | "3" | "4" | "5" | "6" | "7" | "8"
<chiral> ::= "@" | "@@"
<hcount> ::= "H" | "H" <DIGIT>
<charge> ::= "-" | "-" <DIGIT> | "-" <DIGIT> <DIGIT> | "+" | "+" <DIGIT> | "+" <DIGIT> <DIGIT>
<bond> ::= "-" | "=" | "#" | "/" | "\"
<ringbond> ::= <DIGIT> | <bond> <DIGIT>
<branched_atom> ::= <atom> | <atom> <RB> | <atom> <BB> | <atom> <RB> <BB>
<RB> ::= <RB> <ringbond> | <ringbond>
<BB> ::= <BB> <branch> | <branch>
<branch> ::= "(" <chain> ")" | "(" <bond> <chain> ")"
<chain> ::= <branched_atom> | <chain> <branched_atom> | <chain> <bond> <branched_atom>
\end{lstlisting}
\end{tabular}
\caption{\textbf{Context-free grammar used by ChemGE in Backus–Naur form.}}
\label{fig:grammar}
\end{figure*}

\subsection{ChemGE}
\begin{figure}[h!]
\centering
\includegraphics[width=0.7\textwidth]{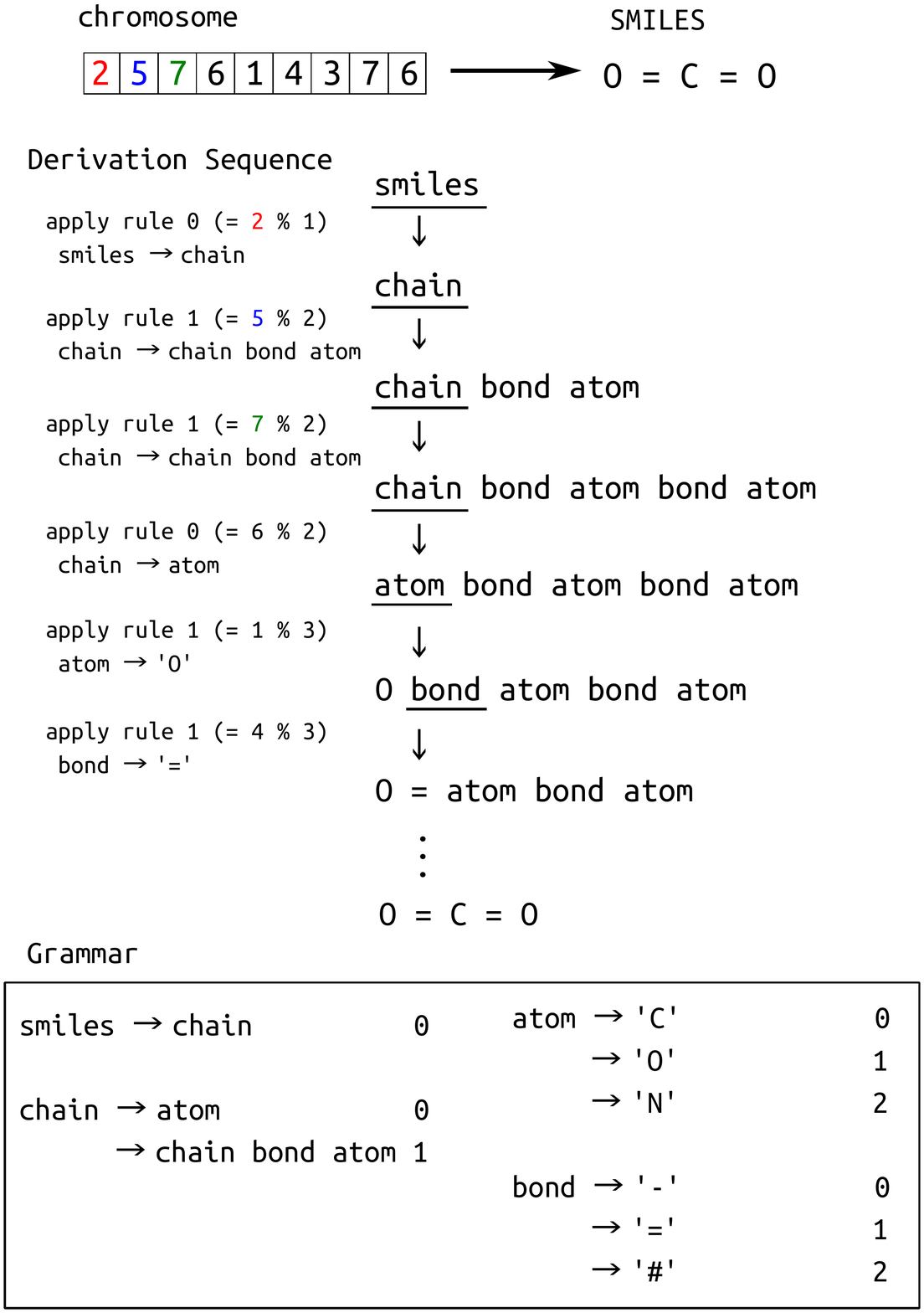}
\caption{\textbf{Mapping process.}
This figure shows how the mapping process translates a chromosome into SMILES.
A chromosome is a sequence of integers. At the $k$-th step, the $k$-th integer is looked up
and it specifies the production rule applied to the leftmost non-terminal symbol.}
\label{fig:mapping}
\end{figure}

ChemGE maintains a population of $\mu$ molecules, each of which is doubly encoded.
A molecule is represented as a SMILES string, which is then encoded as a
sequence of $N$ integers called a {\em chromosome} $C$.
A chromosome is translated to a SMILES string by a mapping process (Figure~\ref{fig:mapping}): the initial string contains the start symbol only, and is successively turned into the final string by
picking up the leftmost non-terminal symbol and applying a production rule to it.
At the $k$-th step of the mapping process,
the $k$-th integer in the chromosome $c = C[k]$ is looked up.
Denote by $r$ the number of rules applicable to the leftmost non-terminal symbol.
Among them, the $((c \mod r) + 1)$-th rule is chosen and applied.
The process iterates until the string has no non-terminal symbol
or the final integer of the chromosome is used.

Evolution is conducted according to the $(\mu+\lambda)$ evolution strategy~\cite{rechenberg2000}
as follows:
(1) create $\lambda$ new chromosomes by repeatedly drawing a random chromosome from the population, and changing one integer at a random position to a random value (i.e., mutation).
(2) Each of the $\lambda$ chromosomes is translated to a SMILES string and then to a molecule and its fitness is evaluated (e.g. by docking simulation). If translation fails, the fitness is set to $-\infty$.
(3) A new generation is made by selecting the $\mu$ top-fit molecules
from the merged pool of existing and new molecules.
We did not use a crossover operation here, as it did not improve the performance (data not shown).

\section{Results and Discussion}
To validate the usefulness of our method, we conducted two types of experiments to generate
1) drug-like molecules, and 2) high scoring ligands for a protein.

\subsection{Benchmark on druglikeness score}
We optimized the $J^{\mathrm{logP}}$ score \cite{gomez2016}, which is an indicator of druglikeness.
It is the octanol-water partition coefficient (logP) penalized
by synthetic accessibility~\cite{ertl2009} and number of carbon rings of size larger than 6.
The score $J^{\mathrm{logP}}$ of a given molecule $m$ is defined as
$$
J^{\mathrm{logP}}(m) = \mathrm{logP}(m) - \mathrm{SA}(m) - \text{ring-penalty}(m) \,,
$$
where $\mathrm{logP}(m)$, $\mathrm{SA}(m)$
and $\text{ring-penalty}(m)$ are normalized so that their mean is zero
and standard deviation is one.
Roughly speaking, molecules with larger $J^{\mathrm{logP}}$ scores have more suitable structural profiles as pharmaceutical drugs.
Since very large logP values ($\mathrm{logP} > 5$)
cause low metabolic stability and other pharmacokinetic defects,
$J^{\mathrm{logP}}$ values of launched drugs and clinical
tested compounds collected from
Clarivate Analytics Cortellis database are ranging from -10 to 5.

We compared ChemGE with CVAE~\cite{gomez2016},
GVAE~\cite{kusner2017}, and ChemTS~\cite{yang2017} using this score.
CVAE uses variational autoencoder (VAE)~\cite{kingma2014} to obtain continuous representation of SMILES strings, conduct Bayesian optimization over the continuous space,
and obtain strings from the optimized representation.
GVAE is an updated version of CVAE,
where grammar information on SMILES is taken into account.
ChemTS uses Monte Carlo tree search and recurrent neural networks
to design SMILES strings.

In this experiment, ChemGE is applied with different initial population sizes: $(\mu,\lambda) = (10,20), (100,200), (1000,2000), (10000,20000)$, randomly chosen from the ZINC database,
which contains 35 million commercially available compounds~\cite{irwin2012}.
$J^{\mathrm{logP}}(m)$ was used as fitness score,
but set to $-\infty$,
if the molecular weight was bigger than 500
or an identical molecule had already been generated before.
In order to compare with other methods, no parallel computation was conducted here.
We used a computing core of Intel Xeon CPU E5-2630 v3.

Table~\ref{tab:logp} shows the maximum score of
each method after running for 2, 4, 6, and 8 hours.
The importance of choosing a suitable population size is well-documented in evolutionary algorithm literature~\cite{dempsey2009},
as an overly small population cannot have sufficient diversity,
while an overly large population cannot optimize each molecule sufffiently due to the small number of generations.
We found that, at $(\mu, \lambda)=(1000, 2000)$,
the final ChemGE score was largest and outperformed other methods in score and speed of generating molecules.
The efficiency of ChemGE enables to evaluate a much larger number of molecules than computationally-demanding deep learning models, such as RNN or VAE, or than Bayesian optimization, which is used in CVAE and GVAE, and gets slower as search progresses.

\begin{table*}[h!]
\centering
\caption{Maximum score $J$ at time points 2, 4, 6, and 8 hours achieved by different molecular generation methods. The average values and standard deviations over 10 trials are shown. The two numbers after ChemGE specifies $(\mu, \lambda)$.  The rightmost column shows the number of generated molecules per minute (duplication is eliminated). }
\label{tab:logp}
\resizebox{\textwidth}{!}{
\begin{tabular}{lllllll}
\hline
Method & 2h & 4h & 6h & 8h & Molecules/Min & Generation \\ \hline
ChemGE $(10,20)$& $4.46\pm 0.34$ & $4.46 \pm 0.34$ & $4.46 \pm 0.34$ & $4.46 \pm 0.34$ & $14.5 \pm 4.0$ & $103000 \pm 24000$ \\
ChemGE $(100,200)$& $5.17\pm 0.26$ & $5.17 \pm 0.26$ & $5.17 \pm 0.26$ & $5.17 \pm 0.26$ & $135 \pm 22$ & $8360 \pm 1340$ \\
ChemGE $(1000, 2000)$& $4.45\pm 0.24$ & $5.32 \pm 0.43$ & $5.73 \pm 0.33$ & $5.88 \pm 0.34$ & $527 \pm 62$ & $704 \pm 59$ \\
ChemGE $(10000, 20000)$& $4.20\pm 0.33$ & $4.28 \pm 0.28$ & $4.40 \pm 0.27$ & $4.53 \pm 0.26$ & $555 \pm 68$ & $72.5 \pm 9.5$  \\
CVAE~\cite{gomez2016} & $-30.18\pm 26.91$ & $-1.39\pm 2.24$ & $-0.61\pm 1.08$ & $-0.006\pm 0.92$ & $0.14\pm 0.08$ & - \\
GVAE~\cite{kusner2017} & $-4.34 \pm 3.14$ & $-1.29\pm 1.67$ & $-0.17\pm 0.96$ &  $0.25\pm 1.31$ &$1.38\pm 0.91$ & - \\
ChemTS~\cite{yang2017} & $4.91\pm 0.38$ &  $5.41\pm 0.51$ & $5.49\pm 0.44$ & $5.58\pm 0.50$ & $40.89\pm 1.57$ & - \\
\hline
\end{tabular}}
\end{table*}

\subsection{Design of high-scoring molecules for thymidine kinase}
ChemGE was applied to the design of high-scoring molecules that are predicted to have strong binding affinity for a specific target protein using rDock~\cite{rdock}:
one of the fastest and most accurate docking simulation program,
commonly used in high throughput virtual screening.
We used thymidine kinase (KITH), a well-known target of antiviral drugs,
as the target protein for this study.
After taking the structure data from a KITH-ligand complex (PDB ID: 2B8T), we generated a cavity using the default reference ligand method (radius: 6 \AA).
To calculate ChemGE's fitness, we took the best intermolecular score ($S_{inter}$) among three rDock runs
from different initial conformations under default parameter settings.
The fitness was defined as $-S_{inter}$, because a small intermolecular score implies high affinity.
If the molecular weight was bigger than 500 or the molecule was already generated, the fitness was set to $-\infty$.

\begin{figure}[H]
\centering
\includegraphics[width=0.8\textwidth]{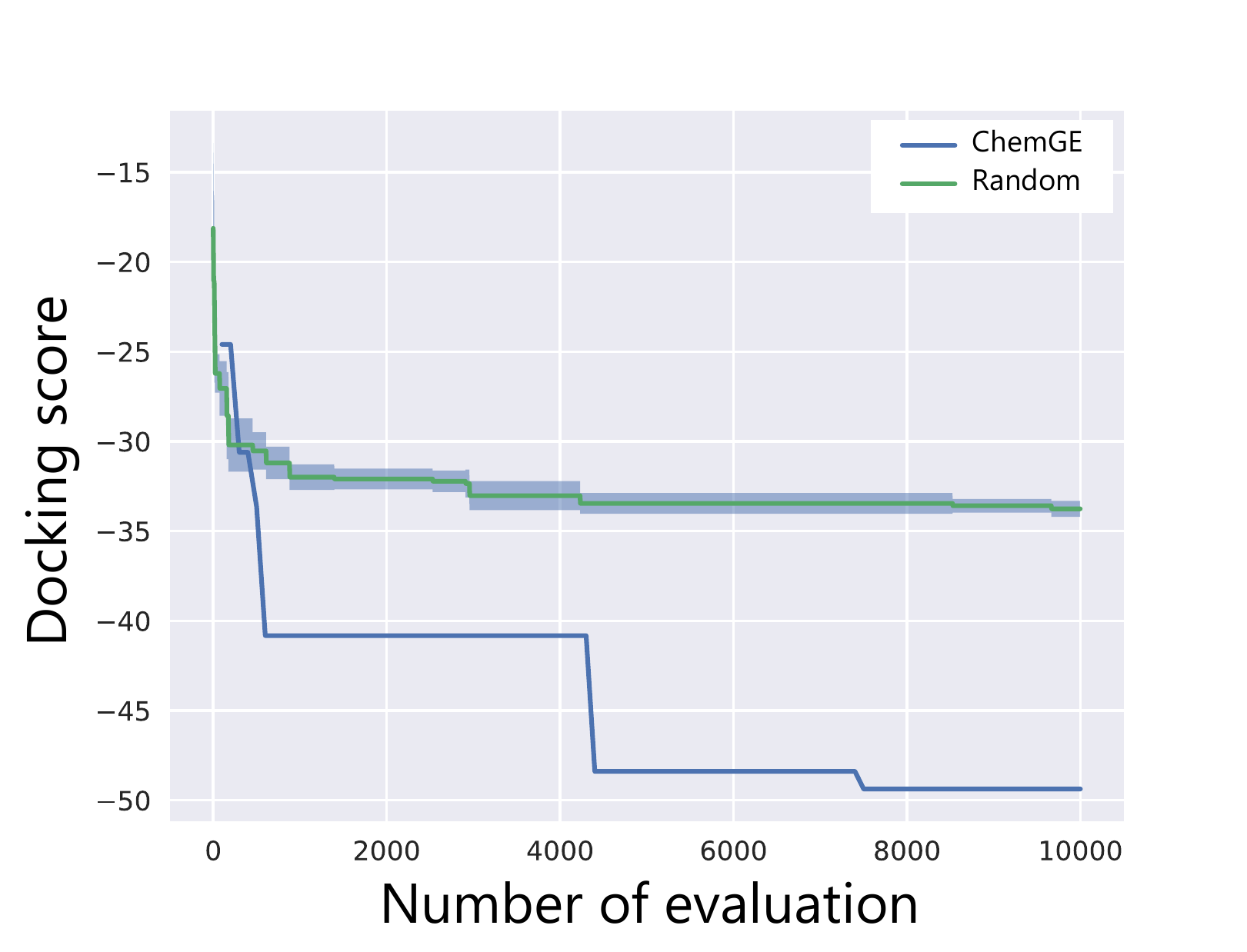}
\caption{\textbf{Progress of molecule optimization by ChemGE and random sampling.} The best intermolecular score computed by rDock and random sampling is shown. For random sampling,
we also showed standard deviation over four trials.}
\label{fig:chemgetest}
\end{figure}

In order to first test its performance, we generated up to 10,000 molecules with ChemGE that were then evaluated with rDock, starting from an initial populations randomly chosen from ZINC.
As can be seen in Figure~\ref{fig:chemgetest}, which shows the progress of the best intermolecular score over number of evaluations, ChemGE optimizes molecules faster than random sampling, by a large margin.

\begin{figure}[H]
\centering
\includegraphics[width=0.8\textwidth]{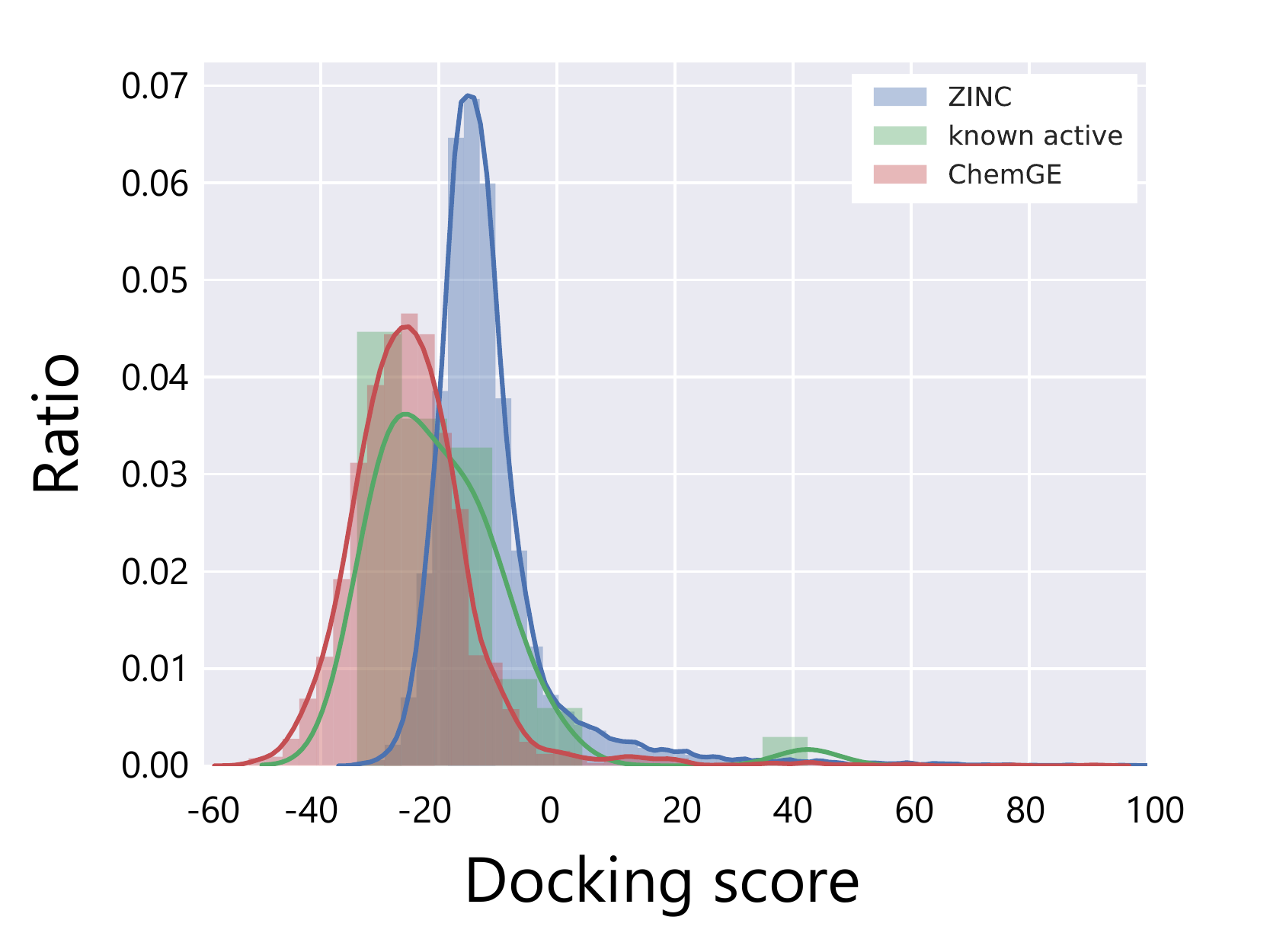}
\caption{\textbf{Score distributions.}  The distribution of rDock's intermolecular score of the molecules generated by ChemGE (red) is biased more favorably in comparison to that of ZINC (green) and that of known thymidine kinase inhibitors in DUD-E (blue).}
\label{fig:rdock-score-dist}
\end{figure}

Next we proceeded to construct the molecular library.
We reserved 32 cores for calculation and set $(\mu, \lambda)=(32, 64)$.
Evaluation of molecules in a population was conducted in parallel.
The computation for 1000 generations was finished in about 26 hours,
resulting in a large library of 9466 generated molecules.
Figure~\ref{fig:rdock-score-dist} shows the distribution of intermolecular scores of ChemGE-generated molecules.
We compared it with the score distribution of 57 known inhibitors derived from DUD-E~\cite{dud-e}, and the baseline score distribution of ZINC.
Knowing that a small intermolecular score implies high affinity, we can see that the distribution of ChemGE is biased towards high affinity in comparison to ZINC, showing the success of evolutionary optimization.
More remarkably, it is also more biased than that of known inhibitors.
We found 349 molecules whose intermolecular score was better than
the best score of 57 known inhibitors. Figures S1-S3 show all of these molecules.
Although synthetic routes and ADMET properties are unknown for these molecules, this large-scale library can offer great opportunities and inspiration for medical chemists.

We investigated the diversity of molecules found by ChemGE.
Larger diversity generally increases the chance of survival at multiple ADMET endpoints.
Using Morgan fingerprints~\cite{rogers2010}, the internal diversity of a set of molecules
$A$ is defined as
\[
I(A) = \frac{1}{|A|^2}\sum_{(x, y) \in A \times A} T_d(x, y),
\]
where $T_d$ is the Tanimoto distance $T_d(x, y) = 1 - \frac{|x \cap y|}{|x \cup y|}$.
We evaluated a "ChemGE-active" molecule set consisting of 349 molecules
whose scores were better then the best known inhibitor.
The internal diversity was 0.55, larger than that of known inhibitors, 0.46.
The difference is substantial, considering that the diversity of the whole ZINC database is 0.65.
Deep reinforcement learning methods often generate very similar molecules~\cite{benhenda2017} and
special coutermeasures are necessary to maintain diversity.
In grammatical evolution, diversity is inherently built-in,
because a large fraction of molecules are mutated in each generation.

\begin{figure}[H]
\centering
\subfloat[Known inhibitors]{\includegraphics[width=0.48\textwidth]{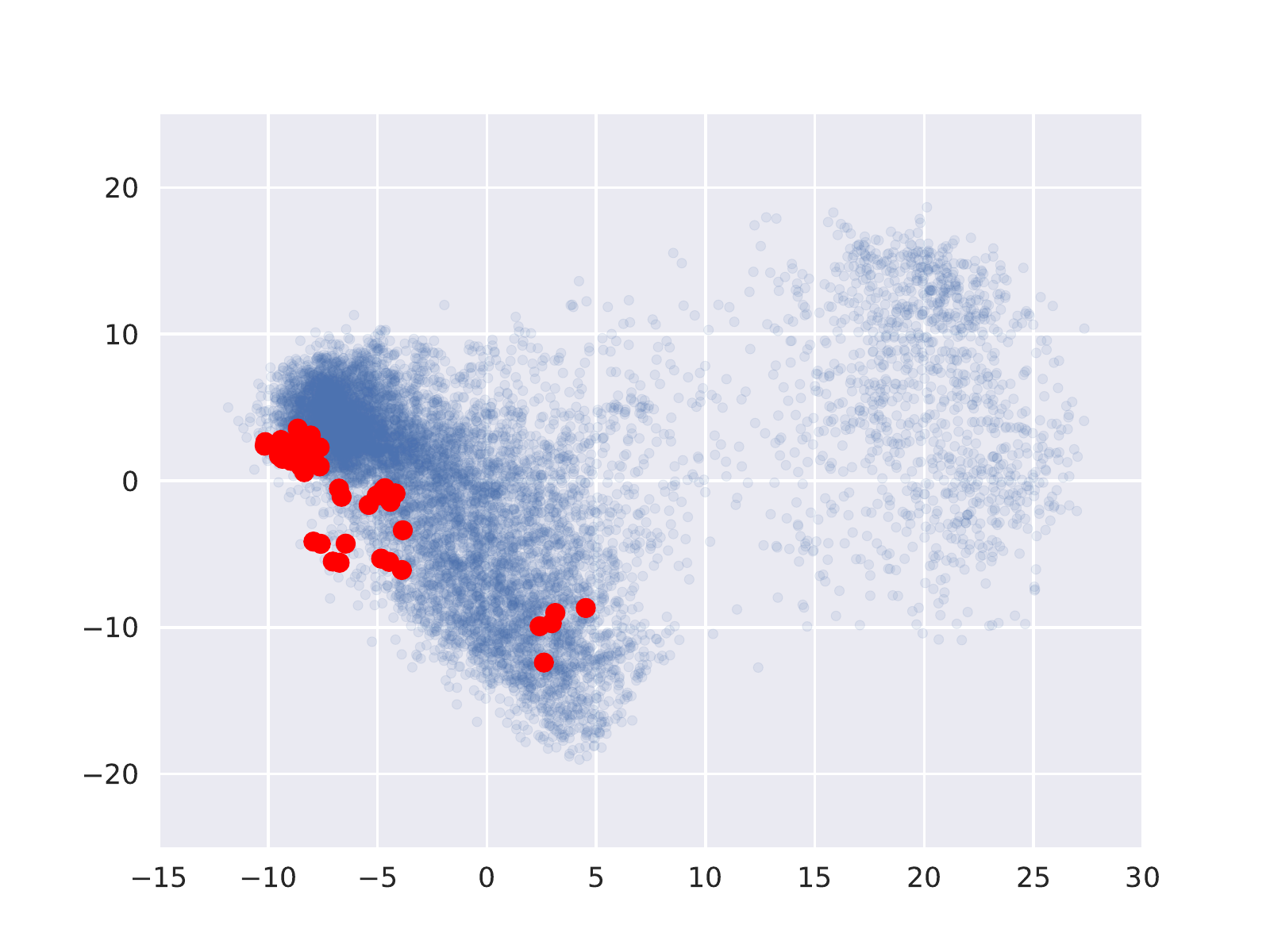}}
\subfloat[Initial population (generation 0)]{\includegraphics[width=0.48\textwidth]{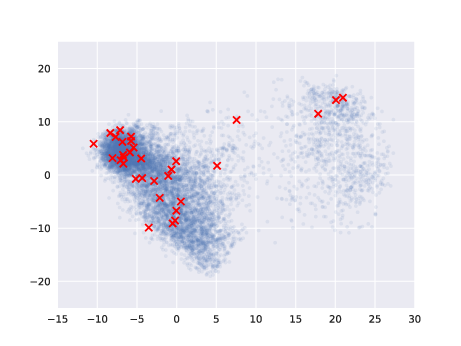}}\\
\subfloat[During exploration (generation 100)]{\includegraphics[width=0.48\textwidth]{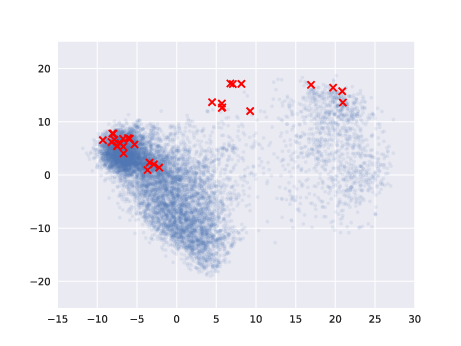}}
\subfloat[Final population (generation 1000)]{\includegraphics[width=0.48\textwidth]{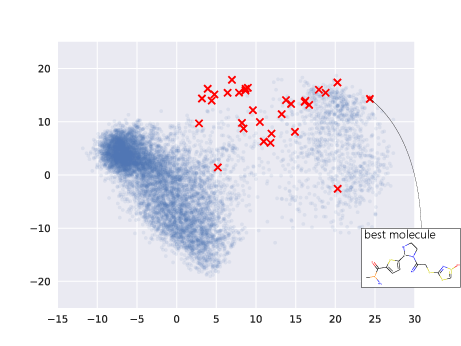}}
\caption{\textbf{ISOMAP visualization of the chemical space.}
Blue dots represent molecules in ZINC database.
Red dots in (a) represent known inhibitors of thymidine kinase.
Red crosses in (b),(c),(d) represent ChemGE populations in different generations.}
\label{fig:rdock-isomap-generation}
\end{figure}

Figure~\ref{fig:rdock-isomap-generation} shows a visualization of
how the chemical space is explored by ChemGE.
A mapping from fingerprints to a two dimensional space is constructed by applying the ISOMAP algorithm~\cite{tenenbaum2000} to ZINC.
Then, the following molecule sets are mapped to the two dimensional space:
(a) known inhibitors, (b) initial population of ChemGE, (c) 100-th generation, (d) 1000-th generation.
The initial population is distributed in various places and a large fraction lies close to known inhibitors. As evolution goes on, the distribution moves away from the known inhibitors
and the final generation occupies a completely different place.
This observation suggests that ChemGE detected a new class of binding molecules,
although they need to be confirmed by biological assays, and
synthesizability and ADMET issues would have to be solved.

\section{Conclusions}
To summarize, we developed ChemGE, a new method to generate functional molecules using grammatical evolution.
In benchmark experiments,
ChemGE efficiently generated far more molecules than deep learning methods, at similar resource levels.
In computational docking experiments with thymidine kinase,
a large library of activity-biased molecules were successfully obtained.
Among them, hundreds of molecules had better scores than known active molecules.
Our library was shown to retain high diversity,
which may contribute to increasing the survival rate in the following steps of the drug discovery process.
Our method is suitable for parallel computation and
our implementation can be applied to even larger computational environments including
cloud platforms such as Amazon Web Services (AWS).
This work demonstrated that molecule generation is possible without costly deep learning and
showed a new direction for research. Nevertheless, to increase performance further,
exploitation of a probablistic or neural model in the evolutionary process
might be beneficial~\cite{ollivier2017information}.

\paragraph{Availability of data and material}
Our implementation is available at \url{https://github.com/tsudalab/ChemGE}.

\paragraph{Competing interests}
The authors declare that they have no competing interests.

\paragraph{Author's contributions}
N. Y., K. O., and K. Tsuda proposed the idea of the molecular design algorithm. N. Y. and K. Terayama implemented the algorithm and performed the experiments. N. Y., K. Tsuda, T. H. and K. Terayama analyzed the experiments. All authors discussed the results and wrote the manuscript.

\paragraph{Acknowledgements}
A part of this work is done when N. Yoshikawa was an intern at Preferred
Networks, Inc. The authors would like to thank David duVerle for
fruitful discussions.
This work is supported by 'Materials research by Information
Integration' Initiative (MI2I) project and Core
Research for Evolutional Science and Technology (CREST)
[grant number JPMJCR1502] from Japan Science and Technology
Agency (JST). In addition,
this work is supported
by Ministry of Education,Culture, Sports, Science and
Technology (MEXT) as 'Priority Issue on Post-K computer'
(Building Innovative Drug Discovery Infrastructure Through
Functional Control of Biomolecular Systems).

\bibliographystyle{unsrt}
\bibliography{reference}

\section*{Supplementary report}
This supplementary report shows the generated molecules whose intermolecular scores are better than known inhibitors.

\begin{figure}
\centering
\includegraphics[width=\textwidth]{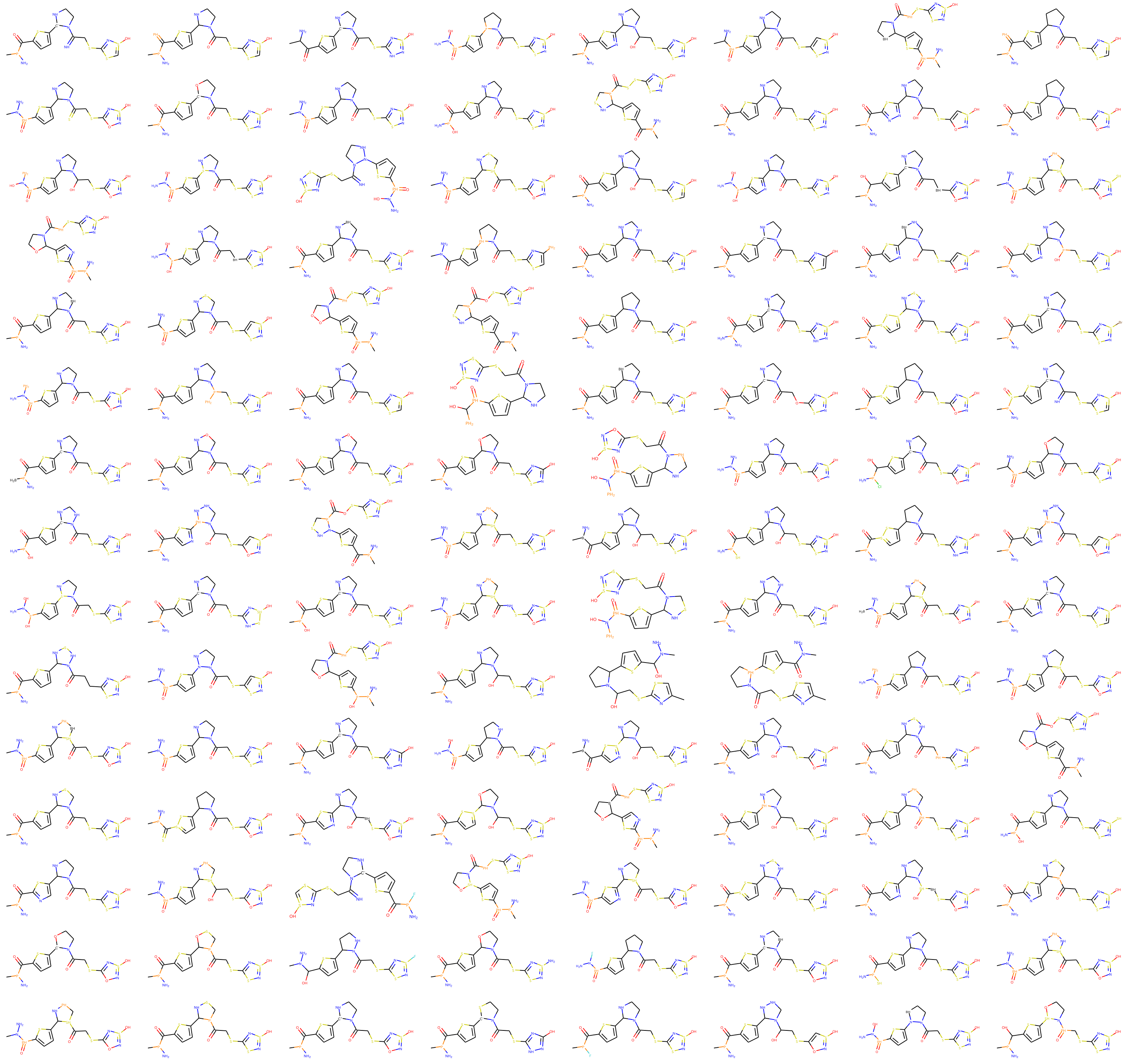}
\caption{All molecules whose scores are better than the best known molecule. Part 1.}
\label{fig:rdock-good-mols-all-1}
\end{figure}
\begin{figure}
\centering
\includegraphics[width=\textwidth]{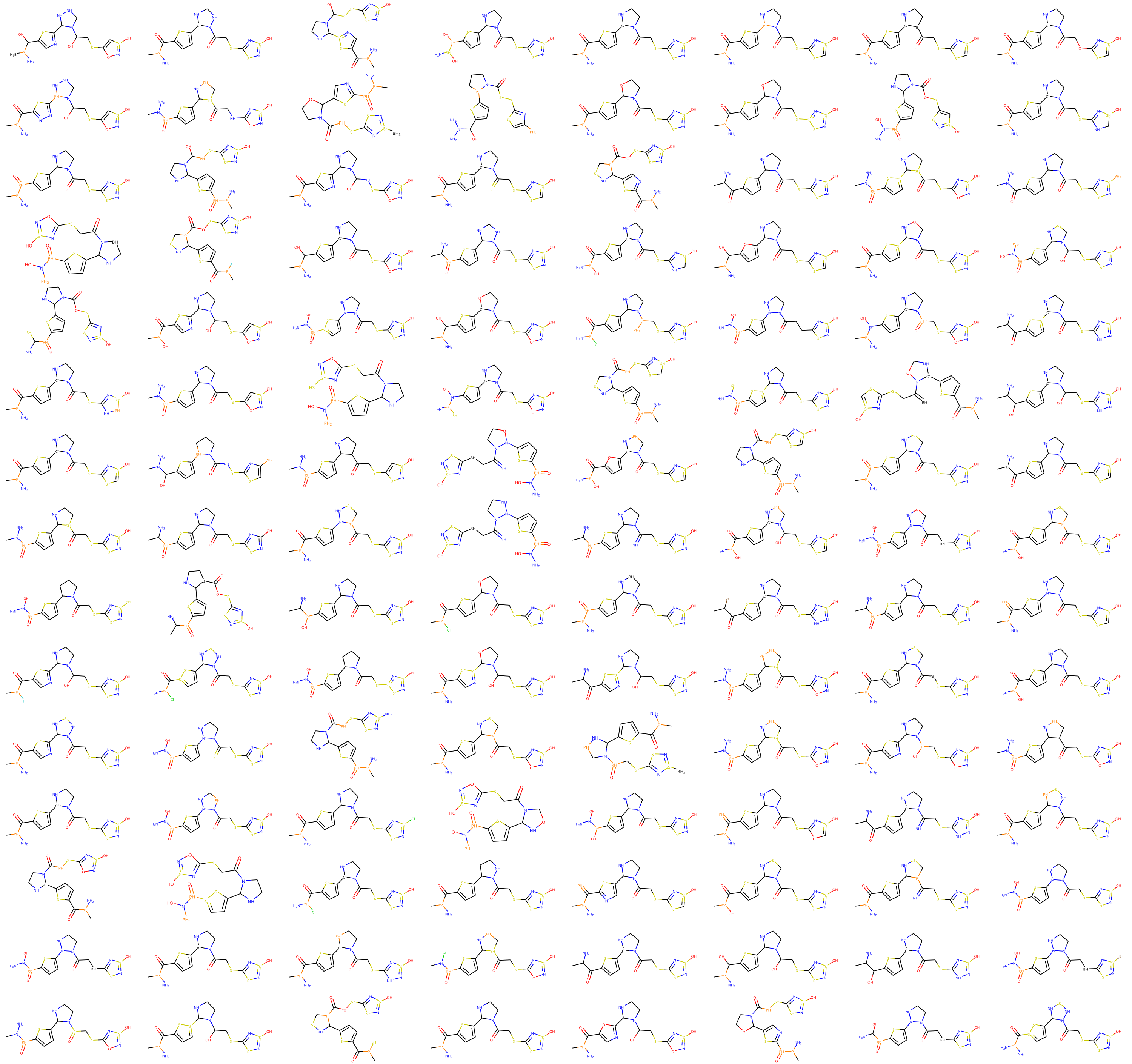}
\caption{All molecules whose scores are better than the best known molecule. Part 2.}
\label{fig:rdock-good-mols-all-2}
\end{figure}
\begin{figure}
\centering
\includegraphics[width=\textwidth]{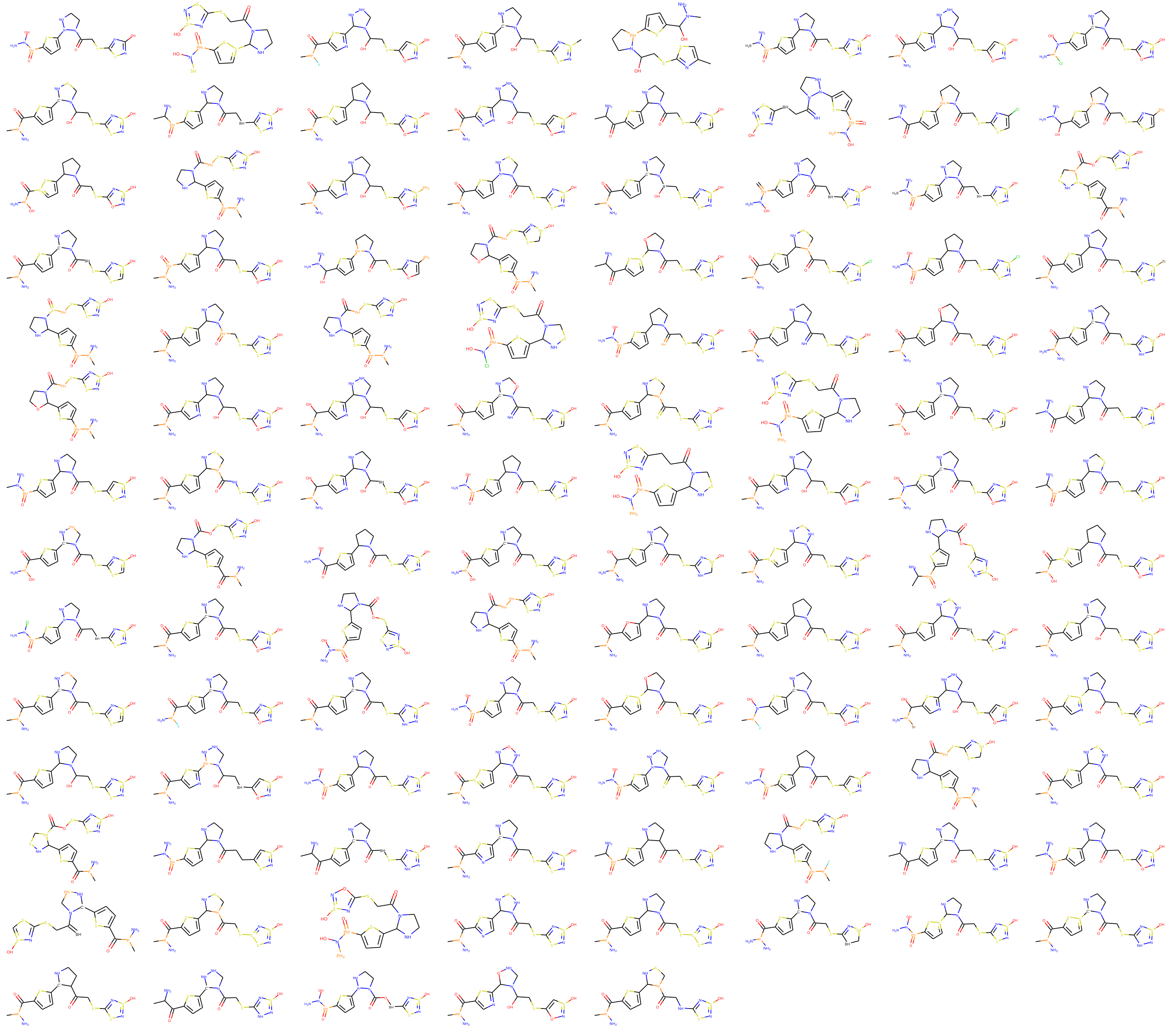}
\caption{All molecules whose scores are better than the best known molecule. Part 3.}
\label{fig:rdock-good-mols-all-3}
\end{figure}

\end{document}